\documentclass[aip,reprint]{revtex4-1}

\draft 

\usepackage{graphicx}
\usepackage{amsmath}
\usepackage{ccaption}
\usepackage{url}
\usepackage{color}

\captionnamefont{\small\normalfont}
\captiontitlefont{\small\normalfont} \captionstyle{\flushleftright}
\captiondelim{.} \normalcaption{} \captionwidth{\linewidth}

\begin{document}

\title{Efficient self-consistent quantum transport simulator for quantum devices}

\affiliation{Sandia National Laboratories, 1515 Eubank SE, Albuquerque, NM 87123, USA}
\author{X. Gao}
\email[]{xngao@sandia.gov}
\author{D. Mamaluy}
\affiliation{Sandia National Laboratories, 1515 Eubank SE, Albuquerque, NM 87123, USA}
\author{E. Nielsen}
\author{R. W. Young}
\affiliation{Sandia National Laboratories, 1515 Eubank SE, Albuquerque, NM 87123, USA}
\author{A. Shirkhorshidian}
\affiliation{Sandia National Laboratories, 1515 Eubank SE, Albuquerque, NM 87123, USA}
\affiliation{University of New Mexico, Albuquerque, NM 87131, USA}
\author{M. P. Lilly}
\author{N. C. Bishop}
\author{M. S. Carroll}
\author{R. P. Muller}
\affiliation{Sandia National Laboratories, 1515 Eubank SE, Albuquerque, NM 87123, USA}

\date{\today}

\begin{abstract}
We present a self-consistent one-dimensional (1D) quantum transport simulator based on the Contact Block Reduction (CBR) method, aiming for very fast and robust transport simulation of 1D quantum devices. Applying the general CBR approach to 1D open systems results in a set of very simple equations that are derived and given in detail for the first time. The charge self-consistency of the coupled CBR-Poisson equations is achieved by using the predictor-corrector iteration scheme with the optional Anderson acceleration. In addition, we introduce a new way to convert an equilibrium electrostatic barrier potential calculated from an external simulator to an effective doping profile, which is then used by the CBR-Poisson code for transport simulation of the barrier under non-zero biases. The code has been applied to simulate the quantum transport in a double barrier structure and across a tunnel barrier in a silicon double quantum dot. Extremely fast self-consistent 1D simulations of the differential conductance across a tunnel barrier in the quantum dot show better qualitative agreement with experiment than non-self-consistent simulations.
\end{abstract}



\maketitle

\section{INTRODUCTION}
\label{sec:intro}

Since the discovery of quantum dots (QDs) in early 1980s \cite{Ekimov1981,Reed1988}, quantum dots have been extensively studied for many applications such as field effect transistors \cite{Hetsch2013}, solar cells \cite{Kamat2013}, light emitting devices \cite{Qasim2013}, and quantum computing \cite{Loss1998}. In the past decade, silicon quantum dot systems have attracted significant interest \cite{Borselli2011,Yamahata2012} exploring their potential use as quantum bits (qubits) for quantum computation, due to their long spin coherence times \cite{Morello2010} and the well-established silicon nanoelectronic manufacturing infrastructure. They are one of the most-studied candidates for qubit application \cite{Nordberg2009,Amir2012}, in part due to the flexibility of forming either spin or charge qubits. A typical experimental double QD (DQD) is described in Sec.~\ref{subsec:experiment}. In most applications of Si QDs, the tunnel barriers' dependence of differential conductance on the external biases represents one of the essential elements of the QD performance. Characterization of the conductance across a tunnel barrier can provide information on the barrier shape and the existence of disorder in the barrier region. This information can, in turn, provide insight on the controllability of the barrier, which is important for manipulating the operation of QDs. However, design and characterization of tunnel barriers in Si QDs, as well as in other material systems, have relied heavily on incremental development through experiment without the presence of efficient computer aided design (CAD) tools for QDs (though they are readily available for classical transistor design and analysis).

Recently, we reported \cite{Gao2012,GaoJAP2013} the development and application of a Quantum Computer Aided Design (QCAD) device simulator that solves for electrostatic potentials, single-particle electronic states through a self-consistent Poisson-Schrodinger solver, and multi-electron states through a Poisson-Schrodinger-Configuration Interaction (P-S-CI) solver. The QCAD simulator (or ``QCAD'' for brevity) is based on the finite element discretization, has a MPI parallel capability, and is applicable to 1D, 2D, and 3D devices. Currently, QCAD has a basic non-self-consistent 1D quantum transport capability that is not sufficient for modeling transport across tunnel barriers in QDs under relatively high drain-source voltages (more than a few $k_B T/q$).

To aid in the understanding of measured transport results and to improve the design of QDs, we need a self-consistent quantum transport modeling capability. Ideally, we would solve a 3D quantum transport problem in a self-consistent way, which would require excessive development time and computing resources, because of the inherent complexity of 3D quantum transport and the sometimes relatively large size of realistic QDs and coupled QD systems of interest. Instead, we propose a very fast, yet qualitatively accurate transport simulation scheme that allows one to analyze and optimize different device geometries and voltages in the ``real-time'' regime. The proposed approach consists of three major steps. First, one performs QCAD 3D \emph{electrostatic} simulations of a given QD device and calibrates threshold voltages to measurement data using gate interface charges. Secondly, with the calibrated interface charges inserted in the device, one runs QCAD 3D simulations for zero drain-source bias and the given experimental gate voltages, and extract the 1D tunnel barrier along the saddle point path determined by a saddle path search algorithm in QCAD. Lastly, one ``feeds'' the extracted barrier to a \emph{self-consistent} 1D quantum transport code to obtain the current-voltage (I-V) characteristics. This three-step procedure allows for very short simulation time and qualitatively accurate transport results (compared to experiment) that provide good general guidance on device designs.

In this paper, we focus on the details of the self-consistent 1D quantum transport code, which is based on adapting the general Contact Block Reduction (CBR) method \cite{Mamaluy2003,Mamaluy2005,Mamaluy2007,Birner2009} to 1D open systems. CBR provides a very efficient technique of implementing the non-equilibrium Green's function (NEGF) formalism \cite{Keldysh1964,DattaQuanTran} for quantum transport, and has been used successfully in simulating 2D and 3D quantum devices \cite{Mamaluy2005,Mamaluy2007,HRyu2012}. Our 1D self-consistent CBR transport highlights three original features. First, we derive and present specific CBR equations optimized for the 1D case for the first time. It turns out that when applying the 3D CBR formulation to 1D applications, we obtain a set of very simple equations that allow for easy implementation and very fast simulation. Secondly, we not only implement the predictor-corrector scheme \cite{Trellakis1997} to achieve the CBR-Poisson self-consistency, but also implement and discuss the scheme with the Anderson acceleration method \cite{Wang2009,Anderson1965}, and demonstrate that the Anderson-accelerated scheme shows superior convergence behaviour for multiple barrier structures. Thirdly, we propose and implement an effective doping technique that allows one to extract the effective doping from a potential profile obtained in an external electrostatic simulator such as QCAD and use it in the self-consistent CBR-Poisson code to simulate transport through the potential.

The reminder of the article is organized as follows. Sec.~\ref{subsec:experiment} briefly describes the relevant experimental setup and results. The self-consistent 1D CBR-Poisson simulator is described in detail in Sec.~\ref{sec:sccbr1dsim}, and its application is demonstrated in Sec.~\ref{sec:app}. Sec.~\ref{sec:conclusion} concludes the article.

\subsection{Experiment}
\label{subsec:experiment}

\begin{figure}
\centering\includegraphics[width=4.2cm, bb=150 280 390 500]{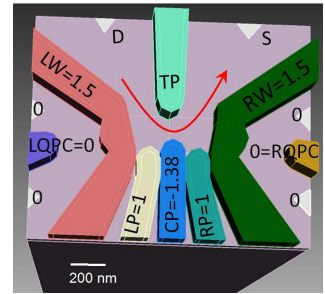}
\caption{\label{fig:DQDStruct} 3D CAD drawing of a lateral double quantum dot structure, showing the depletion gates and silicon substrate. The structure is designed to have two adjacent quantum dots with remote charge sensors located to the left and right of the quantum dots. The voltages indicated in the figure are used to operate the device so that only a single barrier exists between the source, S, and drain, D. The white triangular regions represent Ohmic contacts. The white line denotes a dimension bar.}
\end{figure}

Figure \ref{fig:DQDStruct} shows the 3D CAD drawing of a lateral DQD structure \cite{Tracy2010,Nguyen2013}, showing the depletion gates and silicon substrate. The depletion gates' electrodes are made of highly n-doped polysilicon and rest on a 35-nm SiO$_2$ layer. A global aluminum gate (AG) electrode and a 60-nm Al$_2$O$_3$ insulating layer lie above the polysilicon depletion gates but are not shown in the figure for plot clarity. The AG gate is positively biased to induce electrons in the silicon near the Si/SiO$_2$ interface. The polysilicon gates, TP (top plunger), LP (left plunger), CP (center plunger), RP (right plunger), LW (left wire), and RW (right wire), are biased to produce a single QD, two adjacent QDs, or a single barrier, one example of which is shown in Fig.~\ref{fig:DQDStruct}. The LQPC (left quantum point contact) and RQPC (right quantum point contact) gates can be biased along with LW and RW gates, respectively, to form a narrow conductive constriction, for which the resistance changes appreciably when the total charge in the QDs changes by a single electron. This change in resistance can be used to sense the total charge state of the QDs. Tunnel barriers are formed in a constriction when a sufficiently negative bias is applied to raise the conduction band above the Fermi energy in that region, for example, the dot tunnel barrier between TP and CP. Note that the dot tunnel barrier has a split-gate geometry. As will be shown in Sec.~\ref{subsec:dqdtb}, a split-gate tunnel barrier has a remarkably linear dependence of barrier height on gate voltages.

\begin{figure}
\centering\includegraphics[width=7.0cm, bb=60 210 485 555]{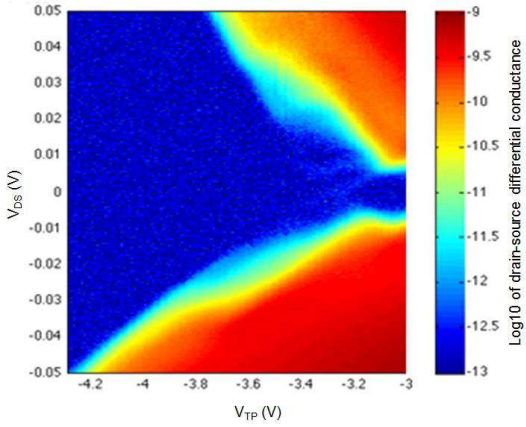}
\caption{\label{fig:ExperimentGds} Measured drain-source differential conductance in logarithmic scale of the dot barrier as a function of $V_{DS}$ and $V_{TP}$ for the DQD device shown in Fig.~\ref{fig:DQDStruct} at T = 4 K. }
\end{figure}

One typical transport measurement of the dot barrier is fixing all the depletion gates and Ohmic contacts at certain voltages, as shown in Fig.~\ref{fig:DQDStruct}, except that the drain-source (DS) and TP voltages are varied. The current flows between the drain and source Ohmic contacts, denoted by the red arrow. The drain-source differential conductance, $G_{DS}=dI/dV_{DS}$, was measured by stepping the TP voltage $V_{TP}$ and sweeping the drain-source voltage $V_{DS}$, and is plotted in logarithmic scale as color contour in Fig.~\ref{fig:ExperimentGds}. The measurement was done using a standard lock-in technique with the sample plunged in liquid Helium (T $\approx$ 4 K), and the measured differential conductance, dI/$dV_{DS}$, is reported in the figure. It is seen that the conductance shows an exponential increase as the barrier height is lowered through increasing $V_{TP}$ (less negative values). As the $V_{TP}$ moves to more negative bias, it requires higher magnitude of $V_{DS}$ to turn on the current conduction, which shows a trade-off between these two voltages' effects on the barrier height and width.

\begin{figure}
\centering\includegraphics[width=7.0cm, bb=65 165 440 670]{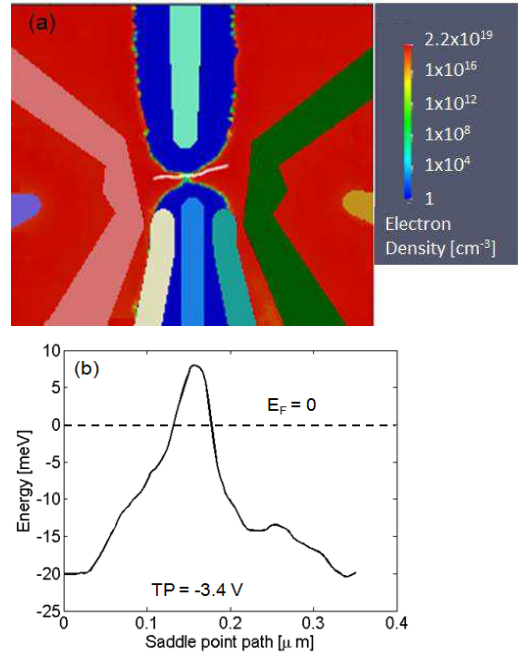}
\caption{ \label{fig:QCADEDensTB} (a) Electron density in silicon near the Si/SiO$_2$ interface obtained from the QCAD 3D simulation under the voltages given in Fig.~\ref{fig:DQDStruct}. The left and right regions are flooded with electrons except at the TP/CP constriction where a potential barrier is formed. The white line denotes the saddle point path determined by a search algorithm in QCAD. The 1D potential energy barrier along the white line is plotted in (b) with the dashed black line indicating the equilibrium Fermi level. }
\end{figure}

For the DQD in Fig.~\ref{fig:DQDStruct} with the given voltages, the resulting electron density from the QCAD simulation is shown in Fig.~\ref{fig:QCADEDensTB}(a), where the left and right regions are flooded with electrons except at the TP/CP constriction where a potential barrier is formed. Although the potential barrier is 3D in nature, we approximate the barrier using a 1D shape along the saddle point path denoted by the white line in the TP/CP constriction. The 1D potential energy barrier at $V_{TP}$ = -3.4 V is plotted in Fig.~\ref{fig:QCADEDensTB}(b). The saddle point and the path are determined using a searching algorithm in QCAD as explained in Ref.~\onlinecite{GaoJAP2013}. The QCAD results here include the effects of gate interface fixed charges, densities of which were obtained by calibrating the thresholds of the device with experimental data.

The measured conductance shown in Fig.~\ref{fig:ExperimentGds} was obtained by sweeping a two-dimensional voltage space, which requires a large number of data points. In order to obtain such a data-intensive plot through simulation in a short time, we apply the self-consistent 1D CBR-Poisson simulator described in Sec.~\ref{sec:sccbr1dsim} to obtain transport data across many 1D potentials similar to Fig.~\ref{fig:QCADEDensTB}(b) under various $V_{TP}$ voltages, and then map the 1D results to approximate 3D through a fitting parameter as explained in Sec.~\ref{subsec:dqdtb}.

\section{SELF-CONSISTENT 1D CBR-POISSON SIMULATOR}
\label{sec:sccbr1dsim}

In this section, details of the 1D CBR formalism are first presented (Sec.~\ref{subsec:cbr1d}), followed by a description of the energy discretization used (Sec.~\ref{subsec:ediscretization}); then details of two self-consistent iteration schemes are explained in Sec.~\ref{subsec:scsolution}, and Sec.~\ref{subsec:Neff} presents how to feed an external potential energy profile to the self-consistent transport code through the effective doping.

\subsection{1D CBR formalism}
\label{subsec:cbr1d}

The non-Hermitian effective Hamiltonian $\mathbf{H}$ of an open quantum device is often written as \cite{DattaQuanTran}
\begin{equation}
\label{eq:effectiveH}
\mathbf{H} = \mathbf{H^0} + \mathbf{\Sigma},
\end{equation}
where $\mathbf{H^0}$ is the Hermitian Hamiltonian of the closed device with \emph{Dirichlet} boundary condition (BC), and $\mathbf{\Sigma}$ is the self-energy. As explained in Ref.~\onlinecite{Mamaluy2003}, within the CBR method, $\mathbf{H}$ can be equivalently written as,
\begin{equation}
\mathbf{H} = \mathbf{H^N} + \mathbf{\Sigma^N},
\end{equation}
where $\mathbf{H}^N$ is the Hamiltonian of the closed device with generalized \emph{Neumann} BC for its eigenfunctions, and $\mathbf{\Sigma}^N$ is the modified self-energy corresponding to $\mathbf{H}^N$. The motivation for using generalized Neumann BC is discussed in Ref.~\onlinecite{Mamaluy2003}. Here we just mention that the traditional Dirichlet BC is in general poorly suited for open-system problems, as it forces wave functions to zero at the device contacts, which is incompatible with the plane-wave-like open-system solutions. Moreover, utilization of generalized Neumann BC allows one to use a dramatically reduced set of eigenstates in the spectral representation of the closed system Green's function \cite{Mamaluy2003}.

When applying the general 3D CBR approach to 1D quantum devices, we obtain a set of very simple equations. We refer readers to Appendix \ref{appendix:1DCBR} for the details of obtaining the set of equations. Here we just provide a recipe to compute the electron density and current:

(1) solve the eigenvalue problem, $\mathbf{H}^N \psi_\alpha = \varepsilon_\alpha \psi_\alpha$, of the closed system with Neumann BC for eigenvectors $\psi_\alpha$'s and eigenenergies $\varepsilon_\alpha$'s;

(2) for every energy $E$, compute

(a) $\Sigma_{jj}(E)$ [Eq.~(\ref{eq:Sigmajj})], $\Gamma_{jj}(E)$ [Eq.~(\ref{eq:Gammajj})], and $G_{ij}^0(E)$ [Eq.~(\ref{eq:Gij0})] for $i=1, \; N$ and $j=1, \; N$;

(b) $G_{1N}^R(E)$ [Eq.~(\ref{eq:G1NR})], $T^{1D}(E)$ [Eq.~\ref{eq:tm}], and $\Lambda(E)$ [Eq.~(\ref{eq:Lambda})];

(c) $x_\alpha(E)$ and $y_\alpha(E)$ for every $\alpha$ using Eq.~(\ref{eq:xyalpha});

(d) $\rho_1(z,E)$, $\rho_N(z,E)$, $f_1^{1D}(E)$, and $f_N^{1D}(E)$ using Eq.~(\ref{eq:rho1N_f1D1N});

(3) integrate over energy to obtain the electron density $n(z)$ using Eq.~(\ref{eq:quantumdensity});

(4) integrate over energy to obtain the current $I^{1D}$ using Eq.~(\ref{eq:Current1D}).

The computational cost for computing the electron density using the above CBR formalism is estimated to be only on the order of $N_{grid} N_E N_{eigen}$, where $N_{grid}$ is the number of space grid points, $N_E$ is the number of energy points, and $N_{eigen}$ is the number of selected eigenstates of $\mathbf{H}^N$ used in the calculation. To see how efficient the CBR formalism is, when compared to the standard matrix inversion approach \cite{DattaQuanTran}, we take the potential barrier in Fig.~\ref{fig:QCADEDensTB} (b) as an example, compute the electron density using the two approaches, and then compare the simulation time. For simplicity, we choose to use commercially available routines: for the CBR eigenvalue problem, we use the LAPACK \emph{DSTEVX} routine because of the real symmetric tridiagonal $\textbf{H}^N$ matrix; for the matrix inversion approach, we use the LAPACK \emph{ZGETRF} and \emph{ZGETRI} routines to obtain the retarded Green's function $G^R$ because of the complex non-Hermitian effective Hamiltonian in Eq.~(\ref{eq:effectiveH}). When using $25 \%$ of the full eigenstates of $\mathbf{H}^N$, i.e., $N_{eigen} = 25 \% \; N_{grid}$, where the $25 \%$ is determined according to the recipe in Sec.~\ref{subsec:ediscretization}, CBR takes less than 1 second to obtain the electron density on a single 2.8 GHz Intel\textregistered  Xeon\textregistered processor, whereas the matrix inversion takes about 250 seconds. Note that the time here is for computing the electron density only \emph{once} and it needs to multiply the number of self-consistent iterations (discussed in Sec.~\ref{subsec:scsolution}) and the number of voltage points to get the total time for one current-voltage curve.

\subsection{Energy discretization}
\label{subsec:ediscretization}

The above-described CBR formulation involves the summation over eigenstates of the closed system [cf. Eq.~(\ref{eq:Gij0})] and the integration over the energy space [cf. Eq.~(\ref{eq:quantumdensity})]. In theory, we should sum over all the eigenstates; however, summing over all the eigenstates is extremely expensive in 2D and 3D devices (though not as much in 1D). Fortunately, due to the use of (generalized) Neumann BC in obtaining eigenstates of the closed system, we can sum only a fraction of all the eigenstates to produce sufficiently accurate results, as has been thoroughly demonstrated for 2D and 3D devices in Refs.~\onlinecite{Mamaluy2005}-\onlinecite{Mamaluy2007}. Here we employ the following recipe to determine the required number of eigenstates for the summation in 1D open systems:

(1) set the minimum energy $E_{min}$ to the smaller of the leads' potential energy, i.e., $E_{min}$ = min($V_1$, $V_N$);

(2) set the maximum energy $E_{max}$ to the larger of the leads' quasi-Fermi energy plus $20 k_B T$, i.e., $E_{max}$ = max($E_{F1}$, $E_{FN}$) + $20 k_B T$;

(3) sum the eigenstates whose energies fall within the interval of ($E_{min}$, $E_{max} + E_{cutoff}$), where $E_{cutoff} = 0.5$ eV often found adequate according to simulation experience.

The choice of $20 k_B T$ is made such that the 1D distribution functions (i.e., $f_1^{1D}$ and $f_N^{1D}$) become negligibly small at energies above $E_{max}$. The number of eigenstates that fall within ($E_{min}$, $E_{max} + E_{cutoff}$) is denoted as $N_{eigen}$. For the double barrier structure discussed in Sec.~\ref{subsec:dbs}, at all the voltages considered, the sufficient $N_{eigen}$ is less than $10\%$ of the total number of eigenstates, which helps significantly reduce the overall simulation time.

It is obvious that the energy integration in Eq.~(\ref{eq:quantumdensity}) needs be done numerically. The lower and upper limits for the integration can be set to $E_{min}$ and $E_{max}$ respectively, as the LDOS goes to zero when $E < E_{min}$ and the 1D distribution functions are negligible when $E > E_{max}$ due to their exponential decay. The choice of the energy grid is very important to obtain accurate integration result. For a single barrier 1D device, a uniform energy grid with a constant energy spacing may be adequate. However, in general 1D devices such as multiple barrier structures, there are often resonant states, and a uniform energy grid cannot sufficiently resolve the contributions around resonant peaks. As described in Ref.~\onlinecite{Mamaluy2007}, a solution to this problem is to employ an adaptive energy grid, making use of the fact that resonant energies are close to eigenenergies of the closed system, and the latter are already computed for the CBR method. To implement the adaptive energy grid, we use three user-adjustable parameters: $dE_{min}$ - the minimum energy spacing, $dE_{max}$ - the maximum energy spacing, and $gf$ - the grid spacing increasing factor. The key steps of the adaptive algorithm are as follows:

(1) Form an energy array including $E_{min}$, $E_{max}$, max($V_1$, $V_N$), and the eigenenergies $\varepsilon_{\alpha}$'s of the closed system that fall into the interval of ($E_{min}$, $E_{max}$), and sort the array.

(2) For each energy interval ($E_i$, $E_{i+1}$) in the energy array, if $E_{i+1} - E_i \le dE_{min}$, bisect the interval and use the bisection point as an energy grid point.

(3) For each energy interval ($E_i$, $E_{i+1}$), if $E_{i+1} - E_i > dE_{min}$, start the geometric progression from both ends and refine the interval from both directions. Specifically, first add two energy points at $\pm dE_{min}/2$ distance away from the ends, then add two more points at an additional $\pm dE_{min} \times gf$ distance (energy spacing = $ dE_{min} \times gf$), then two more points at an additional $\pm dE_{min} \times gf^2$ distance (energy spacing = $ dE_{min} \times gf^2$); continue the procedure until the energy spacing is larger or equal to $dE_{max}$, then use $dE_{max}$ as the energy spacing to add more energy points if needed. There are two criteria to stop the progression. The first one is, if the spacing between the latest two energy points is less than or equal to the recalculated energy spacing for the next progression, stop the progression. The second one is, if the spacing between the latest two energy points is greater than the recalculated energy spacing but smaller than or equal to twice the recalculated energy spacing, bisect the interval and stop the progression.

(4) Repeat steps (2) and (3) until all energy intervals in the energy array have properly-divided grid points.

The adaptive algorithm automatically adjusts the energy grid at each voltage and each CBR-Poisson iteration. Our experience shows that the adaptive energy grid leads to significantly better convergence behavior than a uniform grid with a constant energy spacing due to the reduction of integration error, and it uses many fewer energy points than a uniform grid to obtain a similar result. The adaptive energy grid is made possible because the CBR approach naturally provides eigenenergies of the decoupled system, whereas other implementation methods such as the recursive Green's function do not have such information as an essential part of the method, hence they require extra computation to create a similar adaptive energy grid.

\subsection{Self-consistent solution}
\label{subsec:scsolution}

To obtain the charge self-consistent solution, we need to couple the CBR transport with the Poisson equation (only electrons are considered here)
\begin{equation}
\label{eq:Poisson}
-\nabla \cdot (\epsilon \nabla \varphi) = q (-n + N_D),
\end{equation}
where $\varphi$ is the Hartree potential. $\varphi$ is related to the total potential energy $V$ through the relation of
\begin{equation}
\label{eq:Ec}
V = \Delta E_C - q \varphi,
\end{equation}
where $\Delta E_C$ is the conduction band offset, which is equal to zero in a homogenous material.

\begin{figure}
\centering\includegraphics[width=6.8cm, bb=80 140 460 620]{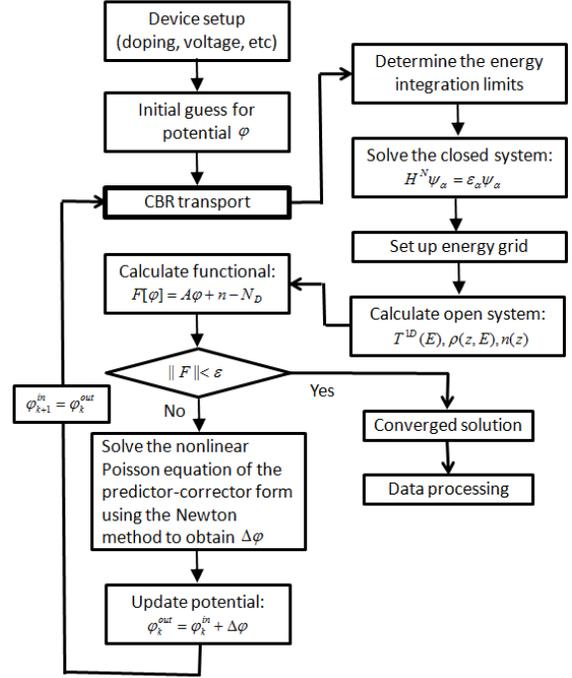}
\caption{\label{fig:CBRPFlowChart} Flow chart of the self-consistent 1D CBR-Poisson transport code.}
\end{figure}

The coupled CBR-Poisson equations can be solved by adapting the predictor-corrector (p-c) iteration scheme \cite{Trellakis1997} to open systems \cite{Mamaluy2007}. Details of the self-consistent p-c procedure are given in Appendix \ref{appendix:selfconsistency} and are summarized in the flow chart of Fig.~\ref{fig:CBRPFlowChart}. It is a simplified version of the general flow chart in Fig. 2 of Ref.~\onlinecite{Mamaluy2007}, specifically adapted to 1D devices. The p-c scheme works very well for most applications even in 3D \cite{Mamaluy2007}, and it typically takes less than 10 iterations to yield a solution with three significant converged (correct) digits in the potential and currents. However, in some applications such as multiple barrier structures, the resonant states make the scheme less efficient, hence we adapted the p-c scheme with the Anderson acceleration \cite{Wang2009,Anderson1965} for open devices.

The main difference between this accelerated scheme and the p-c scheme is that, the $k$th output potential $\varphi_{k}^{out}$ is not directly used as the input potential for the $(k+1)$th iteration. Instead, $\varphi_{k+1}^{in}$ is computed using the Anderson mixing method \cite{Anderson1965}. Namely,
\begin{eqnarray}
\varphi_{k+1}^{in} &=& (1-\beta) \; \bar \varphi_{k}^{in} + \beta \; \bar \varphi_{k}^{out}, \nonumber \\
\bar \varphi_{k}^{in} &=& \varphi_{k}^{in} + \sum_{m=1}^M \theta_k^m (\varphi_{k-m}^{in} - \varphi_{k}^{in} ), \nonumber \\
\bar \varphi_{k}^{out} &=& \varphi_{k}^{out} + \sum_{m=1}^M \theta_k^m (\varphi_{k-m}^{out} - \varphi_{k}^{out} ),
\end{eqnarray}
and the corresponding residual is defined as
\begin{eqnarray}
\bar r_k &=& r_k + \sum_{m=1}^M \theta_k^m (r_{k-m} - r_k), \nonumber \\
r_k &=& \varphi_{k}^{out} - \varphi_{k}^{in}.
\end{eqnarray}
The weighting parameter $\theta_k^m$ is determined by minimizing the inner product of the residual vector
\begin{equation}
\frac {\partial \langle \bar r_k , \bar r_k \rangle} {\partial \theta_k^m} = 0,
\end{equation}
which leads to a linear equation of the form $Ax=b$ with $A_{mn}=\langle r_k - r_{k-m}, r_k - r_{k-n} \rangle$, $b_m = \langle r_k, r_k - r_{k-m} \rangle$, and $x_m = \theta_k^m$. One can see that the Anderson mixing method has two tunable parameters $\beta$ and $M$, which are discussed in more details in Sec.~\ref{subsec:dbs}. The flow chart for the Anderson accelerated scheme is the same as in Fig.~\ref{fig:CBRPFlowChart}, except that the $\varphi^{in}_{k+1} = \varphi_{k}^{out}$ block is replaced by $\varphi_{k+1}^{in} = (1-\beta) \; \bar \varphi_{k}^{in} + \beta \; \bar \varphi_{k}^{out}$.

\subsection{Effective doping}
\label{subsec:Neff}

In a standard device simulation, a doping profile is usually given, and the self-consistent quantum transport code starts the simulation using the given doping profile. When a potential energy profile comes from an external electronic structure simulator such as QCAD, there is no doping information. The question is how one can utilize such a potential at thermal equilibrium to obtain the charge self-consistent transport solution across the potential under an applied bias. The strategy that we propose to address this question is described in the following and is applicable to 1D/2D/3D applications.

\begin{figure}
\centering\includegraphics[width=5.0cm, bb=115 235 375 550]{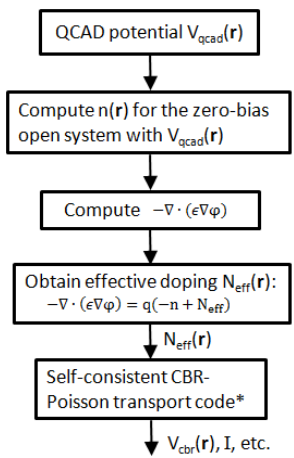}
\caption{\label{fig:NeffFlowChart} Flow chart of obtaining the effective doping $N_\text{eff}(\textbf{r})$ for the self-consistent CBR-Poisson transport code, given the external potential energy profile $V_\text{qcad}(\textbf{r})$. The * indicates that the block corresponds to the flow chart of  Fig.~\ref{fig:CBRPFlowChart}. }
\end{figure}

\begin{figure}
\centering\includegraphics[width=7.0cm, bb=75 215 510 560]{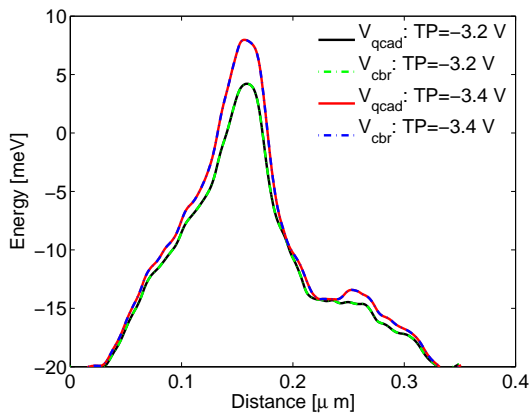}
\caption{\label{fig:VcbrAt0Bias} Comparison of the QCAD potential energy $V_\text{qcad}(\textbf{r})$ and the self-consistent potential energy $V_\text{cbr}(\textbf{r})$ from the CBR-Poisson transport code at zero bias for two different profiles. We see that  $V_\text{cbr}(\textbf{r})$ and $V_\text{qcad}(\textbf{r})$ curves are on top of each other as expected. }
\end{figure}

Given the potential energy profile $V_\text{qcad}(\textbf{r})$, we first apply the open-system CBR formalism under zero bias to compute the quantum electron density $n(\textbf{r})$. (For 1D, use the formulation described in Sec.~\ref{subsec:cbr1d}, and for 2D/3D cases, use the general CBR formulation in Ref.~\onlinecite{Mamaluy2007}.) Then we compute the left hand side of the Poisson equation, $-\nabla \cdot (\epsilon \nabla \varphi)$, for the given $V_\text{qcad}(\textbf{r})$. If the transport occurs in a homogeneous material, $\varphi$ can be chosen as the negative of $V_\text{qcad}(\textbf{r})$ divided by q, i.e., $V_\text{qcad}(\textbf{r}) = -q \varphi$. If the transport occurs in a heterostructure device, we need to compute $\varphi$ using the relation in Eq.~(\ref{eq:Ec}). When computing the second derivative, it is recommended using at least the five-point finite difference method to obtain a smooth result. Next, we solve the Poisson equation for the doping $N_D(\textbf{r})$. As this doping is not physical, but a purely mathematical result, we name it effective doping $N_\text{eff}(\textbf{r})$. Once this effective doping is computed, we can input it to the self-consistent CBR-Poisson transport code shown in Fig.~\ref{fig:CBRPFlowChart}, which allows for obtaining currents at various voltages. The procedure is summarized in Fig.~\ref{fig:NeffFlowChart}. Note that only the effective doping $N_\text{eff}(\textbf{r})$ is passed into the transport code, not the QCAD potential energy $V_\text{qcad}(\textbf{r})$. At zero bias, with the computed $N_\text{eff}(\textbf{r})$, $V_\text{qcad}(\textbf{r})$ simultaneously satisfies the Poisson and the open-system CBR (i.e., Schrodinger) equations, hence the self-consistent result $V_\text{cbr}(\textbf{r})$, obtained from the CBR-Poisson code which uses $N_\text{eff}(\textbf{r})$ as the doping profile and zero as the potential initial guess, is numerically the same as $V_\text{qcad}(\textbf{r})$, as illustrated in Fig.~\ref{fig:VcbrAt0Bias}. Now, for a given $N_\text{eff}(\textbf{r})$, we can obtain charge self-consistent solutions for non-zero biases.

\section{APPLICATION AND DISCUSSION}
\label{sec:app}

The above-described self-consistent 1D CBR-Poisson code has been applied to simulate the quantum transport in a double barrier structure and across the tunnel barrier in a silicon lateral double quantum dot.

\subsection{Double barrier structure}
\label{subsec:dbs}

The double barrier structure (DBS) we considered consists of 20-nm GaAs source region / 5-nm AlGaAs barrier / 5-nm GaAs well / 5-nm AlGaAs barrier / 20-nm GaAs drain region, similar to the structure in Ref.~\onlinecite{Sollner1983}. The source and drain regions are doped with n-type doping of $10^{18}$ cm$^{-3}$, the n-type doping in the well is $10^{17}$ cm$^{-3}$, and the barriers have no intentional dopants. For simulation purpose, we put $10^6$ cm$^{-3}$ (close to the intrinsic concentration of GaAs at room temperature) n-doping in the barriers. The conduction band offset is taken as 0.23 eV, approximately corresponding to the aluminum content of $25\%$ to $30\%$ \cite{Sollner1983}. For simplicity, we use an electron effective mass of $0.063 m_0$, where $m_0$ is the electron free mass, and a relative dielectric constant of 12.9 throughout the entire structure. A lattice temperature of 25 K is assumed, and a uniform grid size of $a$ = 0.2 nm is used.

\begin{figure}
\centering\includegraphics[width=8.0cm, bb=60 215 560 575] {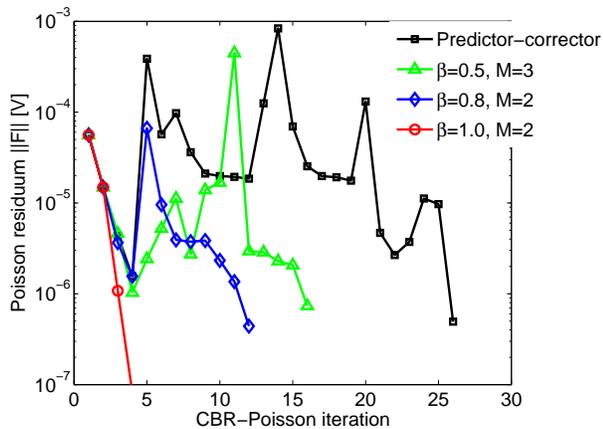}
\caption{\label{fig:GaAsRTDConv0.02V} Convergence comparison of the CBR-Poisson procedure between the p-c scheme and the Anderson-accelerated p-c scheme at 0.02 V. The black curve is obtained using the p-c method alone, while other three curves are obtained using the Anderson acceleration with different $\beta$ and $M$ values. }
\end{figure}

\begin{figure}
\centering\includegraphics[width=8.0cm, bb=85 215 515 555] {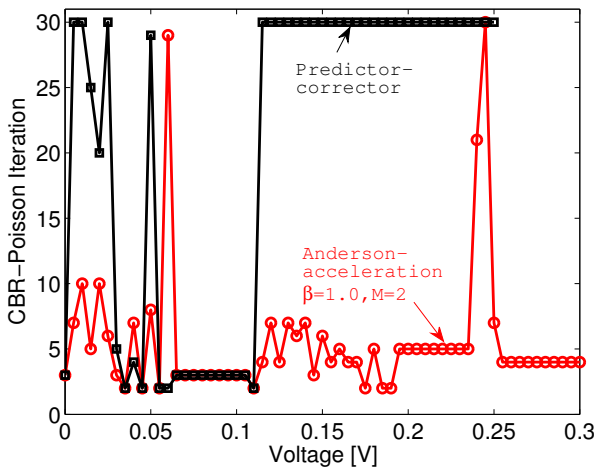}
\caption{\label{fig:GaAsRTDConv} Number of self-consistent CBR-Poisson iterations required to achieve the Poisson residuum $||F|| < 10^{-6}$ V at various voltages using the predictor-corrector method without (black) and with (red) the Anderson acceleration using $\beta=1.0, M=2$.}
\end{figure}

In this particular DBS, the Anderson-accelerated p-c scheme leads to better convergence than that without the acceleration. Figure \ref{fig:GaAsRTDConv0.02V} shows a comparison of the CBR-Poisson convergence behavior between the p-c methods without and with the Anderson acceleration. Each curve is obtained by starting the self-consistent CBR-Poisson simulation with zero potential as the initial guess. The simulation is taken as converged when the Poisson residuum $||F|| < 10^{-6}$ V. It is also worthy of noting that the convergence behavior is very non-monotonic, typical of open systems, unlike the monotonic or nearly monotonic convergence found in closed systems \cite{GaoJAP2013,Trellakis1997,Wang2009}. For the applied voltage of 0.02 V, the selected three sets of $\beta$ and $M$ values for the Anderson scheme all lead to faster convergence. However, there appears no rigorous rule for determining the $\beta$ and $M$ values, which makes it difficult to choose their optimal values. Furthermore, the optimal values may differ for each system and each voltage of interest. (The three sets of values were chosen manually by trials.) It happens that, for the simulated DBS, the Anderson acceleration with $\beta=1.0, M=2$ results in an overall better convergence than the p-c scheme alone for all the voltages considered, as shown in Fig.~\ref{fig:GaAsRTDConv}. The maximum number of CBR-Poisson iterations were limited to 30. It is seen that the p-c scheme alone could not achieve convergence for voltages greater than 0.11 V, while the Anderson-accelerated scheme is able to achieve convergence for all voltages except one voltage point. In general, we expect the p-c scheme with Anderson acceleration would achieve better convergence for quantum devices that contain resonant states such as double or multiple barrier structures, provided one can find the appropriate $\beta$ and $M$ values. Due to the lack of a rigorous rule to determine the optimal Anderson parameters, a better acceleration scheme is still worthy of research. The simulation at each non-zero voltage in Fig.~\ref{fig:GaAsRTDConv} is started using the potential solution obtained at a previous voltage as initial guess, hence the number of CBR-Poisson iterations taken at 0.02 V is somewhat different from the results in Fig.~\ref{fig:GaAsRTDConv0.02V}.

\begin{figure}
\centering\includegraphics[width=8.0cm, bb=55 270 480 635] {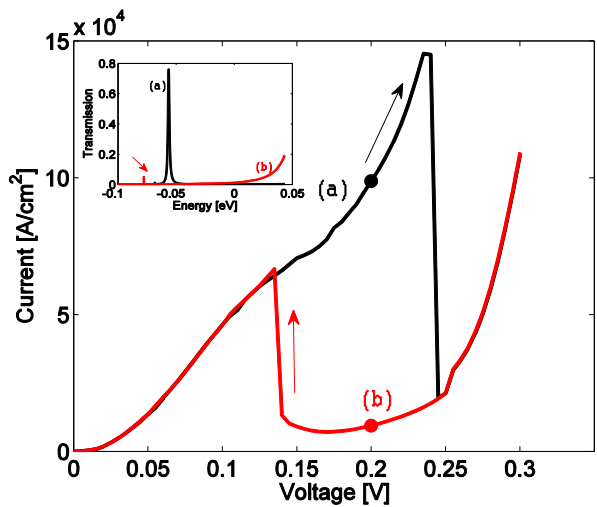}
\caption{\label{fig:GaAsRTDIV} Current-voltage characteristics of a GaAs/AlGaAs double barrier structure. The arrows indicate the direction of the voltage sweeping. The black curve is obtained by increasing the voltage from 0 to 0.3 V with a voltage step of +5 mV, while the red one is obtained by decreasing the voltage from 0.3 V to 0 with a voltage step of -5 mV. Inset shows the transmission energy spectrum for the two points marked as (a) and (b) on the I-V curves. }
\end{figure}

The current-voltage characteristics for the DBS obtained from the self-consistent CBR-Poisson transport code is given in Fig.~\ref{fig:GaAsRTDIV}, which shows clear hysteresis, as observed in numerous articles \cite{Alves1988,Laux2004,Zou1994}. The hysteresis range in this device appears to be quite large, which could be in part due to the neglect of the exchange and correlation potential that has been shown \cite{Zou1994} to reduce the range of bistability in resonant tunneling devices (RTD). It is also noted that we did not intend to simulate realistic RTDs here, but to demonstrate the capability of our transport code on idealized double barrier structures, hence the proper treatment of scattering and lead injection is not considered in this work. One can refer to the work by Klimeck \emph{et al.} \cite{Klimeck1995,Bowen1997} for a more realistic treatment of scattering and lead injection in RTDs. Each I-V curve in Fig.~\ref{fig:GaAsRTDIV} has 61 voltage points and takes about 30 seconds total on a single 2.8 GHz Intel\textregistered Xeon\textregistered processor, which leads to a computation time of about 0.5 second per bias point.

\begin{figure}
\centering\includegraphics[width=8.0cm, bb=70 215 550 560] {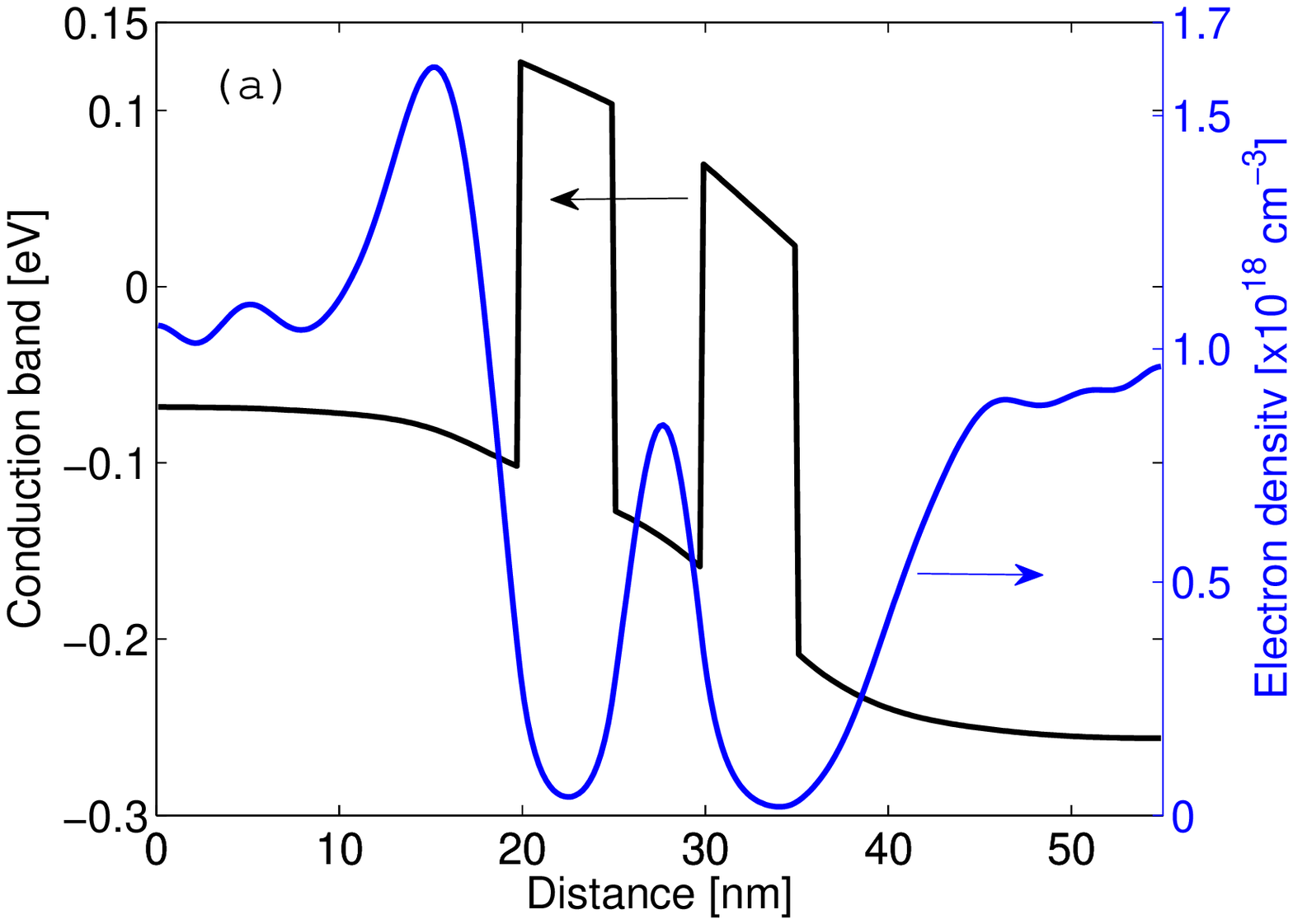}
\centering\includegraphics[width=8.0cm, bb=70 215 550 570] {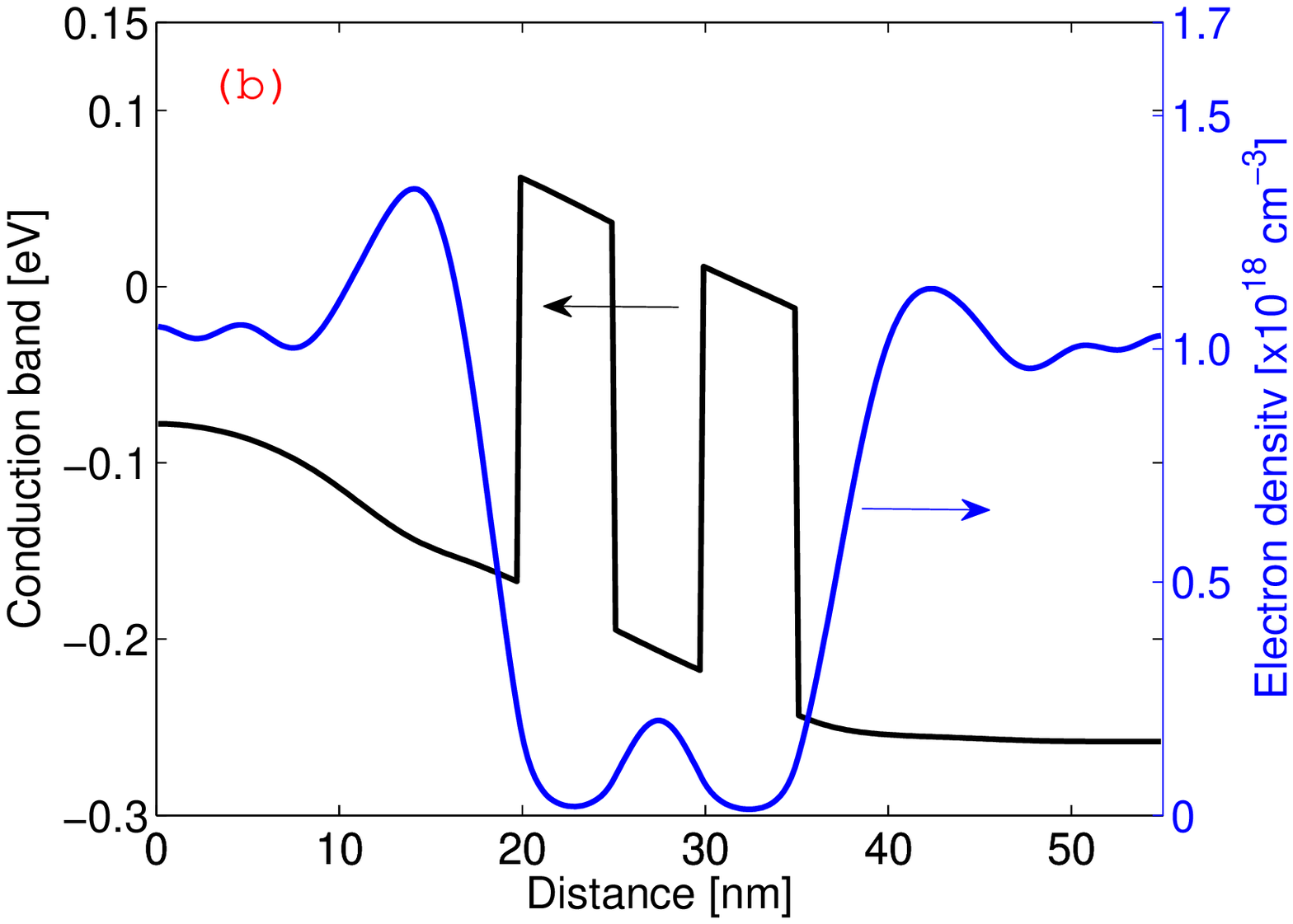}
\caption{\label{fig:EDensCBAt0.2V} Conduction band and electron density vs distance corresponding to the two points marked in Fig.~\ref{fig:GaAsRTDIV} as (a) and (b). The applied voltage is 0.2 V where the current is multivalued, as seen in Fig.~\ref{fig:GaAsRTDIV}, and hence the conduction band and electron density are different for the same applied bias.}
\end{figure}

The conduction band and electron density corresponding to the voltage points marked as (a) and (b) in Fig.~\ref{fig:GaAsRTDIV} are shown in Figs.~\ref{fig:EDensCBAt0.2V}(a)-(b) respectively. In Fig.~\ref{fig:EDensCBAt0.2V}(a), resonant conduction through the ground state of the well is observed, since the electron density displays a single large hump as expected for the ground state solution in the well, and the resonance conduction leads to higher current. In Fig.~\ref{fig:EDensCBAt0.2V}(b), although the electron density also shows a single hump, its magnitude is much smaller, so the device is off resonance, resulting in small current. Note that since there is no significant charge in the well, the potential drop in the well appears linear, and the potential drops in the two barriers are also equal because of the constant electric field in the well. On the other hand, in Fig.~\ref{fig:EDensCBAt0.2V}(a), there is a visible band bending in the well, consistent with the significant electron charge present there, and the right barrier has more potential drop than the left barrier due to the band bending. The resonant conduction at point (a) and the off-resonance at point (b) are also confirmed by the transmission function energy spectrum shown in the inset of Fig.~\ref{fig:GaAsRTDIV}, where curve (a) shows a clear resonant peak whereas curve (b) has no resonance-like peak in transmission. The asymmetric peak marked by the red arrow in curve (b) corresponds to the peak in the local density of states at the conduction band edge of the left lead, not due to resonance.

Note that in Fig.~\ref{fig:EDensCBAt0.2V}, the equilibrium Fermi level at zero voltage is set to 0, the left lead is fixed at 0 V, and the voltage is applied to the right lead. By using the Neumann BC for the Poisson equation as described in Sec.~\ref{sec:sccbr1dsim}, the equilibrium Fermi energy, which is the difference between the equilibrium Fermi level and the conduction band level, is found \emph{automatically} by the self-consistent CBR-Poisson transport code. However, in the case of using the Dirichlet BC, one has to calculate the Fermi energy beforehand, and the calculation almost always relies on certain approximations, for example, using an approximate expression for the inverse of Fermi-Dirac integral.

\subsection{Quantum dot tunnel barrier}
\label{subsec:dqdtb}

As motivated in Sec.~\ref{sec:intro}, it was of experimental importance to be able to model the tunneling current across tunnel barriers in multi-quantum dots. In this section, we apply the self-consistent 1D CBR-Poisson transport code to simulate the current-voltage relation across the dot tunnel barrier in the lateral DQD structure, which is operated to produce only a single barrier at the TP/CP constriction as given in Fig.~\ref{fig:DQDStruct}, under various TP gate voltages at 4 K. We then extract the drain-source differential conductances (i.e., $G_{DS}=dI/dV_{DS}$) and compare them with experiment. As an additional comparison, we also carry out another set of simulations using a non-self-consistent version of the CBR code, which assumes a linear potential energy drop in the barrier region above the equilibrium Fermi level. The electron effective mass along the transport direction is $0.19 m_0$, and the in-plane effective mass is $m_{\parallel}^* = m_0 \sqrt{0.98 \times 0.19}$. The predictor-corrector scheme without the Anderson acceleration is used since it is sufficient to achieve convergence for the single barriers.

\begin{figure}
\centering\includegraphics[width=8.0cm, bb=50 135 485 635] {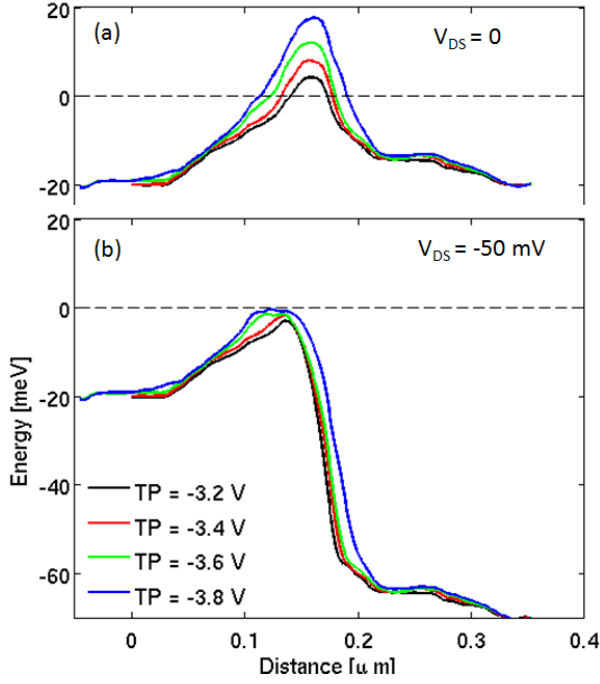}
\caption{\label{fig:CBRP4TPs} (a) Four different tunnel barriers at $V_{DS}$ = 0 obtained from QCAD. (b) Tunnel barriers at V$_{\text{DS}}$ = -50 mV obtained from the self-consistent CBR-Poisson code. The dashed black lines denote the equilibrium Fermi level at 0 (i.e., $E_F = 0$).}
\end{figure}

\begin{figure}
\centering\includegraphics[width=8.0cm, bb=105 230 515 540] {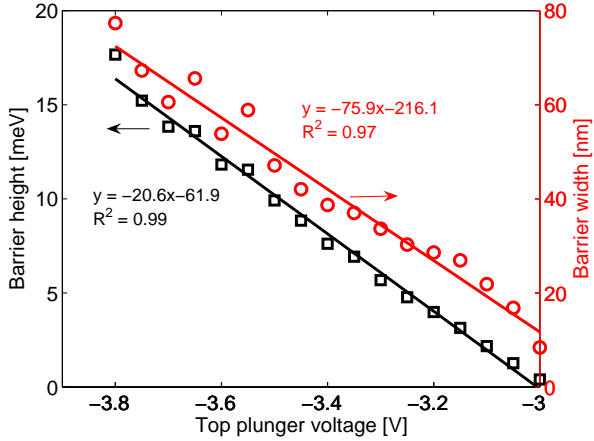}
\caption{\label{fig:VtpBHBW} Barrier height and width as a function of $V_{TP}$. The barrier height is the peak of a potential barrier with respect to $E_F = 0$, and the barrier width is also defined at $E_F = 0$. Symbols are simulated data from QCAD, and lines are linear fits with the fitting equations and least square errors shown. }
\end{figure}

Figure \ref{fig:CBRP4TPs}(a) shows four different tunnel barriers obtained from QCAD by applying different voltages to the TP gate of the DQD device given in Fig.~\ref{fig:DQDStruct}, while keeping other gate voltages unchanged and $V_{DS}$ = 0. The peaks of the barriers and their widths at $E_F = 0$ as a function of $V_{TP}$ are plotted in Fig.~\ref{fig:VtpBHBW}. It is seen that the barrier height shows a nearly ideal linear dependence on the TP voltage. This linear dependence has been asserted in previous simplistic models such as the one by K. MacLean \emph{et al.} \cite{MacLean2007}. Here our 3D QCAD simulation results confirm that split-gate tunnel barriers indeed have barrier heights that depend linearly on the applied voltage. The functional linear dependence of the barrier height on the voltage can be used for phenomenologically 1D modeling of tunnel barriers \cite{AmirInPrep}.

For each barrier in Fig.~\ref{fig:CBRP4TPs}(a), an effective doping profile is extracted according to Sec.~\ref{subsec:Neff}, and is then input to the self-consistent CBR-Poisson code to simulate the transport at different $V_{DS}$ voltages. The self-consistent potential barriers at $V_{DS}$ = 0 obtained from the CBR-Poisson code are numerically the same as the ones obtained in QCAD. And the self-consistent tunnel barriers at $V_{DS}$ = -50 mV obtained from the CBR-Poisson code are shown in Fig.~\ref{fig:CBRP4TPs}(b). We see that, the -50 mV bias is large for the tunnel barriers under consideration, and it significantly changes the barriers' shape and height.

\begin{figure}
\centering\includegraphics[width=8.0cm, bb=75 215 515 580] {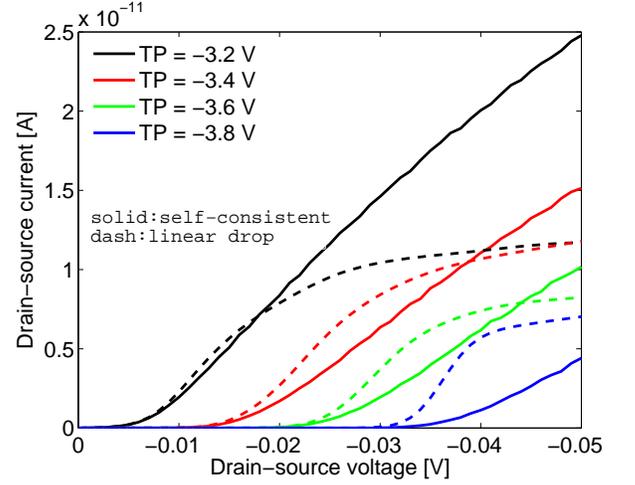}
\caption{\label{fig:CBRPIV} Simulated current-voltage relation for the four tunnel barriers in Fig.~\ref{fig:CBRP4TPs}. The solid curves are obtained from the self-consistent CBR-Poisson code, while the dashed curves are obtained from the non-self-consistent CBR code assuming a linear potential drop in the barrier region above $E_F$.}
\end{figure}

The simulated current-voltage curves for the four tunnel barriers in Fig.~\ref{fig:CBRP4TPs} are given in Fig.~\ref{fig:CBRPIV}. Note the drain-source current on the vertical axis is in units of Amperes. The transformation from the 1D current in units of Amperes/cm$^2$ to the 3D current in Amperes is explained in Appendix \ref{appendix:3Dcurrent}. The drain-source currents are computed using Eq.~(\ref{eq:approxI3D}) with $\gamma = 10^{-4}$. (The determination of $\gamma = 10^{-4}$ is explained below.) The dashed curves in Fig.~\ref{fig:CBRPIV} are obtained from the non-self-consistent CBR code assuming a linear potential drop in the barrier region above $E_F$. Note that the linear potential drop region is wider for a higher barrier such as the red barrier at TP = -3.4 V in Fig.~\ref{fig:CBRP4TPs}(a), than a lower barrier such as the black curve at TP = -3.2 V. It is clear that the dashed curves significantly deviate from the self-consistent currents at high drain-source voltages.

\begin{figure}
\centering\includegraphics[width=8.0cm, bb=75 215 510 560] {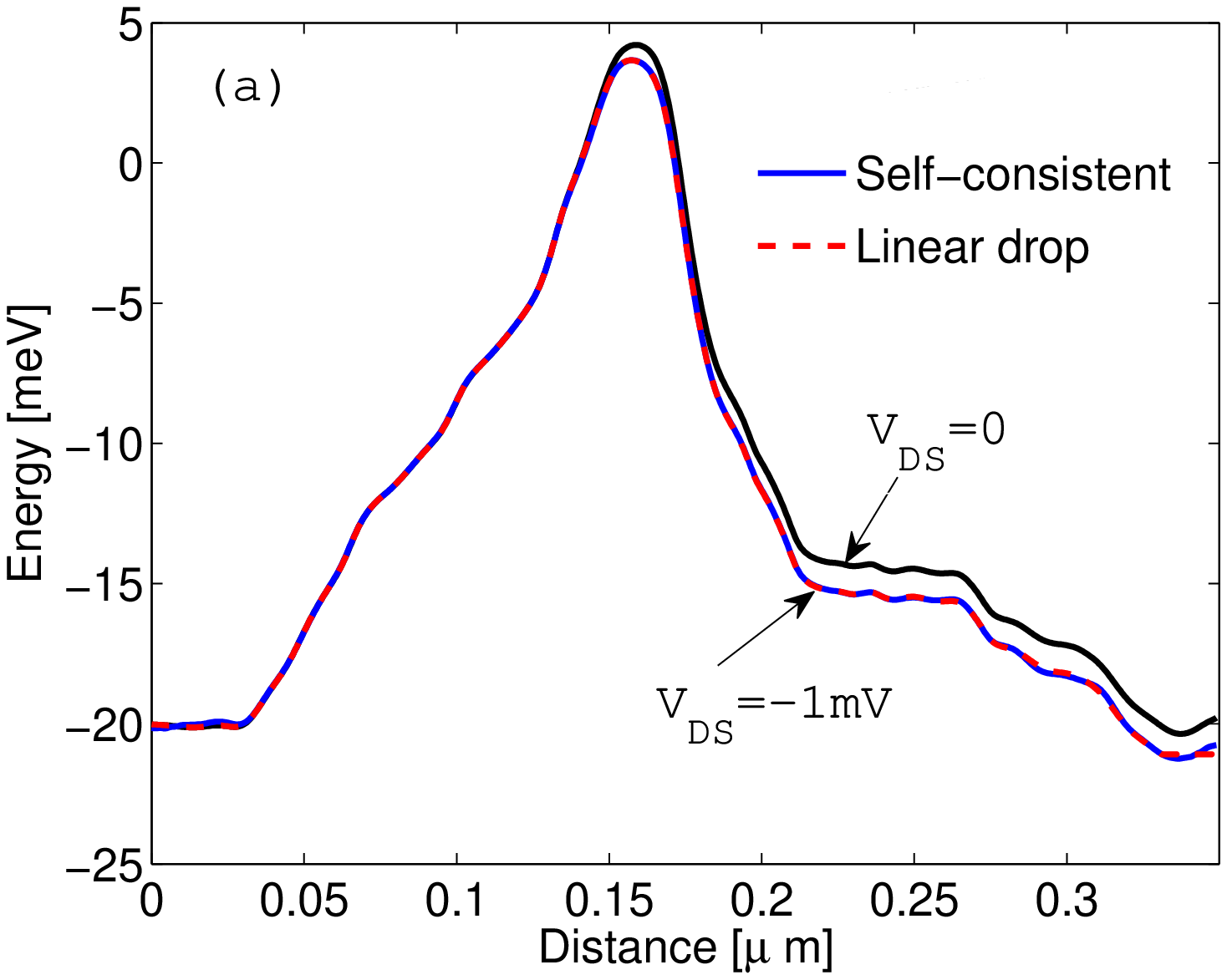}
\centering\includegraphics[width=8.0cm, bb=75 215 510 570] {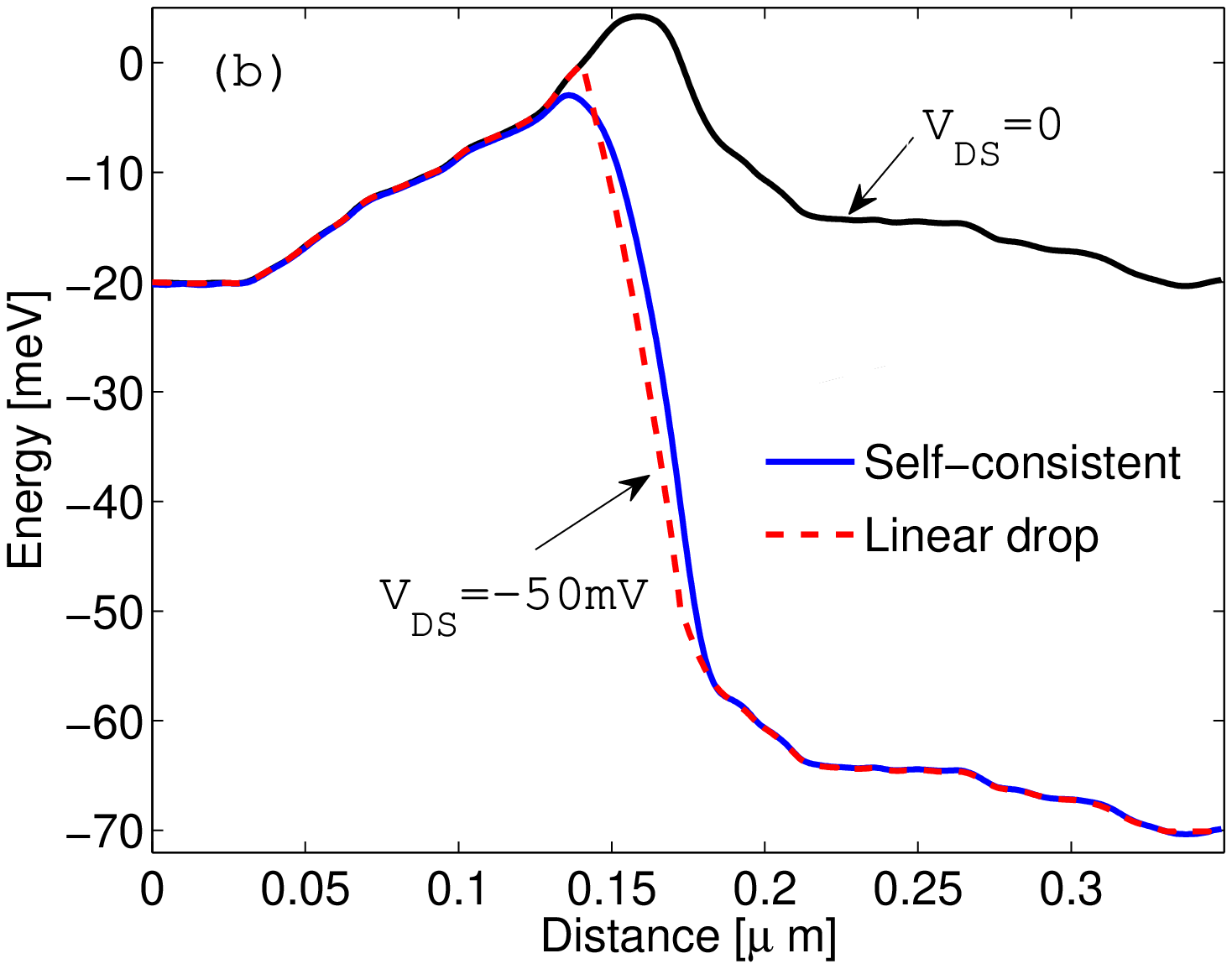}
\caption{\label{fig:CmprsnOfTPs} Potential energy profiles under $V_{TP}$ = -3.2 V at $V_{DS}$ = -1 mV (a) and at $V_{DS}$ = -50 mV (b). The black curve is the barrier at $V_{DS}$ = 0. The blue solid curve is the potential obtained from the self-consistent CBR-Poisson code, whereas the red dashed curve is obtained from the non-self-consistent CBR code assuming a linear potential drop.  }
\end{figure}

To compare the results in more details, we plot the potential energy profiles at $V_{DS}$ = -1 mV and $V_{DS}$ = -50 mV in Fig.~\ref{fig:CmprsnOfTPs}. In (a), the non-self-consistent potential with a linear drop is nearly on top of the self-consistent potential, suggesting that the linear potential drop assumption is adequate at this low voltage, which is a few times of $k_B T/q$ ($k_B T \approx 0.35$ meV at T = 4 K). However, as the $V_{DS}$ increases, the non-self-consistent potential becomes more deviated from the self-consistent solution, as clearly seen in (b), where the red curve shows a small barrier at the source side due to the assumption of the linear potential drop that pins the barrier peak at 0. This artifact leads to the saturation in the non-self-consistent currents and consequently the large differences between the solid and dashed curves in Fig.~\ref{fig:CBRPIV} at high bias.

The experimental data available for comparison with simulation are $G_{DS}$ of the dot tunnel barrier as a function of $V_{DS}$ and $V_{TP}$. The measured differential conductance is shown as color contour in Fig.~\ref{fig:ExperimentGds}. The voltage steps taken for the measurement were very small, e.g., the $V_{TP}$ voltage step is on the order of 10 mV and the $V_{DS}$ voltage step is less than 1 mV, leading to very smooth $G_{DS}$ contour. For simulation, we first obtain the tunnel barriers from QCAD at $V_{DS} = 0$ and various $V_{TP}$ with a voltage step of 50 mV; then for the barrier at a given $V_{TP}$, we run the transport code to obtain the I-V curve with a $V_{DS}$ voltage step of 5 mV, as exemplified by the curves in Fig.~\ref{fig:CBRPIV}. For every I-V curve, we use the finite center difference method to compute $G_{DS}$, and then plot the conductance in color contour as a function of $V_{TP}$ and $V_{DS}$, similar to the plotting of the measured data. To map the calculated maximum $G_{DS}$ to roughly the same order of magnitude as the measured $G_{DS}$, we use a constant fitting parameter $\gamma$ introduced in Eq.~(\ref{eq:approxI3D}) and obtain an approximate value of $\gamma = 10^{-4}$.

\begin{figure}
\centering\includegraphics[width=8.0cm, bb=50 75 425 705] {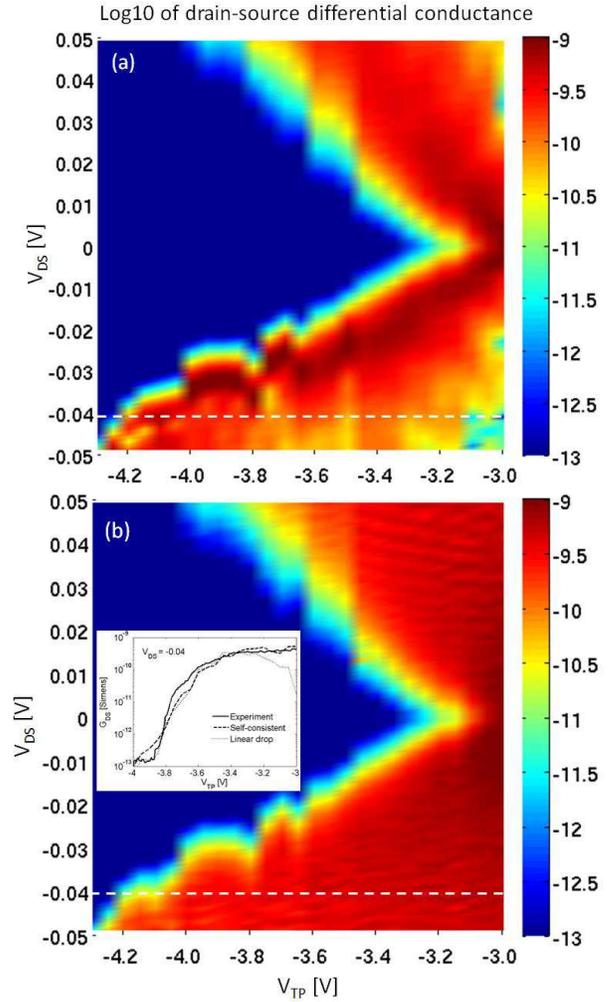}
\caption{\label{fig:CmprsnGsm} Color contour of $G_{DS}$ in logarithmic scale as a function of $V_{DS}$ and $V_{TP}$. (a) Simulated result obtained from the non-self-consistent CBR code assuming a linear potential drop in the barrier region above $E_F$. (b) Simulated result obtained from the self-consistent CBR-Poisson code. The measured differential conductance is shown in Fig.~\ref{fig:ExperimentGds}. Inset in (b) shows a comparison of conductance line cuts at $V_{DS}=-0.04$ V between experiment and simulations. }
\end{figure}

Comparing the results in Fig.~\ref{fig:CmprsnGsm} with the experiment in Fig.~\ref{fig:ExperimentGds}, we observe that the simulated $G_{DS}$ from the self-consistent CBR-Poisson code [Fig.~\ref{fig:CmprsnGsm}(b)] is in much better qualitative agreement with the measured $G_{DS}$ than that of using the non-self-consistent CBR code [Fig.~\ref{fig:CmprsnGsm}(a)], although both sets of simulations are able to capture the asymmetry of $G_{DS}$ with respect to $V_{DS}$. Specifically, the self-consistent $G_{DS}$ shows a color contour map with respect to $V_{DS}$ and $V_{TP}$, similar to that of the measured $G_{DS}$, whereas the non-self-consistent $G_{DS}$ color contour deviates significantly from the measurement, especially in the small $|V_{TP}|$ and high $|V_{DS}|$ region, where the conductance shows very wrong or nonphysical (i.e., negative) values due to the artificial current saturation shown in Fig.~\ref{fig:CBRPIV}. The good agreement between self-consistent conductance and experiment, and the wrong non-self-consistent conductance for more positive $V_{TP}$ is further demonstrated by the inset plot in Fig.~\ref{fig:CmprsnGsm}(b). This observation indicates that non-self-consistent simulations are not sufficient for simulating transport across the tunnel barriers, though they are much faster (due to the absence of self-consistent iterations) and so appealing for use.

The main obvious difference between the measurement [Fig.~\ref{fig:ExperimentGds}] and the self-consistent $G_{DS}$ [Fig.~\ref{fig:CmprsnGsm}(b)] is in the region of small $|V_{DS}|$, where a high resistance persists for more positive $V_{TP}$ than is predicted by the simulations. This high resistance funnel has also been observed in similar experimental devices and appears therefore to be a feature of the split-gate tunnel barrier geometry. In the type of QD devices considered, the effective length (at $E_F$) of the TP/CP constriction perpendicular to the current flow is anticipated to change with voltages due to quantum confinement along the axis. The effective height of the 1D barrier along the current flow direction depends on the changes in this perpendicular dimension because of the inherent coupling of the two directions. The 2D interplay could lead to the high resistance region in the measurement as discussed in Ref.~\onlinecite{AmirInPrep}. This 2D effect is not captured by the present simulations since they use 1D potential profiles. Nevertheless, the self-consistent 1D treatment reproduces the barrier behavior for a large range of $|V_{DS}|$, showing that the approach can produce rapid numeric results consistent with experiment, and providing insight about how the barrier and the transport respond in elevated electric field regions.

\section{CONCLUSION}
\label{sec:conclusion}

We have presented a charge-self-consistent 1D quantum simulator that couples the CBR-based transport with the Poisson equation self-consistently, and aims for fast and robust transport simulations of 1D quantum devices. Specific CBR equations optimized for the 1D case are derived in detail for the first time. The predictor-corrector scheme with the optional Anderson acceleration is implemented to achieve the CBR-Poisson self-consistency. It is found that the Anderson-accelerated scheme shows a superior convergence rate for a double barrier structure. The proposed effective doping strategy allows to extract an effective doping from a potential energy profile obtained in an external electrostatic simulator such as QCAD under thermal equilibrium, and then use the doping in the CBR-Poisson code to obtain the transport solution through the potential under an applied bias. Simulated drain-source differential conductances using the self-consistent code across a tunnel barrier in a silicon QD device show much better qualitative agreement with experimental data than the non-self-consistent model assuming a linear potential drop.

\begin{acknowledgements}
This work was supported by the Laboratory Directed Research and Development program at Sandia National Laboratories. Sandia is a multiprogram laboratory operated by Sandia Corporation, a Lockheed Martin Company, for the United States Department of Energy under Contract No. DE-AC04-94AL85000. The related experiment work was performed at the Center for Integrated Nanotechnologies, a U.S. DOE, Office of Basic Energy Sciences user facility.
\end{acknowledgements}

\appendix

\section{1D CBR Formalism}
\label{appendix:1DCBR}

When applying the general 3D CBR approach to 1D quantum devices for a single band, the generalized Neumann BC is reduced to the standard Neumann condition for the eigenvectors of $\mathbf{H^N}$, i.e., $\psi(1) = \psi(0)$ and $\psi(N) = \psi(N+1)$ .

The modified contact self-energy $\mathbf{\Sigma}_C^N$ is a two-by-two matrix in 1D, and it has just two non-zero elements corresponding to the left and right device boundaries, which are in contact with the external leads denoted as lead $1$ and $N$ respectively,
\begin{equation}
\label{eq:selfenergy}
\mathbf{\Sigma}_C^N = \left[ \begin{array} {cc}
\Sigma_{11} & 0 \\
0 & \Sigma_{NN}
\end{array} \right]
= \left[ \begin{array} {cc}
t_0 -t_0 e^{i k_1 a} & 0 \\
0 & t_0 -t_0 e^{i k_N a}
\end{array} \right],
\end{equation}
where $k_1$ and $k_N$ are the wave vectors in the left and right leads respectively, and $t_0 = \hbar^2 / (2 m^* a^2)$ with $a$ being the uniform grid size and $m^*$ being the electron effective mass along the transport direction. They are determined by the dispersion relation \cite{DattaQuanTran}
\begin{eqnarray}
\label{eq:dispersion}
E &=& V_1 + 2 t_0 [1-\cos(k_1 a)], \nonumber \\
E &=& V_N + 2 t_0 [1-\cos(k_N a)],
\end{eqnarray}
where $V_1$ and $V_N$ are the constant potential energies in the left and right leads respectively. Using Eqs.~(\ref{eq:selfenergy})-(\ref{eq:dispersion}) and the Euler formula, we get
\begin{equation}
\label{eq:Sigmajj}
\Sigma_{jj} = \frac{E-V_j}{2} \pm i \sqrt{ (E-V_j) \biggl( t_0 - \frac{E-V_j}{4} \biggl) },
\end{equation}
with $j = 1, N$ and $i$ preceding the imaginary part.

The Hermitian matrix $\mathbf{\Gamma}_C$ is defined as
\begin{equation}
\label{eq:GammaC}
\mathbf{\Gamma}_C = i (\mathbf{\Sigma}_C^N - \mathbf{\Sigma}_C^{N+}) =
\left[ \begin{array} {cc}
\Gamma_{11} & 0 \\
0 & \Gamma_{NN}
\end{array} \right],
\end{equation}
and the non-zero elements of $\Gamma_{jj} \; (j = 1, N)$ are
\begin{equation}
\label{eq:Gammajj}
\Gamma_{jj} = \mp 2 \sqrt{(E - V_j) \biggl( t_0 - \frac{E - V_j} {4} \biggl) }.
\end{equation}
Since $\Gamma_{jj}$ are often interpreted as ``broadening'' functions \cite{DattaQuanTran}, we choose to use the positive sign of $\Gamma_{jj}$ without loss of generality, which corresponds to the minus sign in Eq.~(\ref{eq:Sigmajj}).

The 1D transmission function between lead $1$ and $N$ is given by \cite{Mamaluy2005}
\begin{eqnarray}
\label{eq:tm}
T^{1D}(E) &=& \mbox{Tr} ( \mathbf{\Gamma}_C^{(1)} \textbf{G}_C^R \mathbf{\Gamma}_C^{(N)} \textbf{G}_C^{R+} )
= \Gamma_{11} \Gamma_{NN} |G_{1N}^R|^2, \nonumber \\
\end{eqnarray}
where \mbox{Tr} means taking the trace, and $\textbf{G}_C^R$ is a two-by-two submatrix of the open device's retarded Green's function within the contact regions. $\textbf{G}_C^R$ is given by $\textbf{G}_C^R = (\textbf{I} - \textbf{G}_C^0 \mathbf{\Sigma}_C)^{-1} \textbf{G}_C^0$, with $\textbf{G}_C^0$ being the submatrix of the closed device Green's function in the contact regions. Making use of the two-by-two matrix algebra, we obtain
\begin{equation}
\label{eq:G1NR}
G_{1N}^R = \frac{G_{1N}^0} {(1-\Sigma_{11} G_{11}^0)(1-\Sigma_{NN} G_{NN}^0) - \Sigma_{11} \Sigma_{NN} |G_{1N}^0|^2 }.
\end{equation}

The elements of $\mathbf{G}_C^0$ are determined through the spectral representation \cite{Mamaluy2005}
\begin{equation}
\label{eq:Gij0}
G_{ij}^0 = \sum_\alpha \frac{\psi_\alpha^*(i) \psi_\alpha (j)} {E - \varepsilon_\alpha}, \; \; i=1,N, \; \; j=1,N,
\end{equation}
with $\psi_\alpha$ and $\varepsilon_\alpha$ being the eigenvectors and eigenenergies of the closed system Hamiltonian with Neumann BC, i.e., $\mathbf{H}^N \psi_\alpha = \varepsilon_\alpha \psi_\alpha$. Note that $\psi_\alpha$ here is normalized according to the Dirac vector normalization condition (not the integration normalization in position space),
\begin{equation}
\label{eq:norm}
\sum_{\alpha, i} \sum_{\beta, j} \psi_\alpha^*(i) \psi_\beta(j) = \delta_{\alpha \beta} \delta_{ij},
\end{equation}
with $\delta$ being the Kronecker delta.

Let $\textbf{B}_C = \textbf{I}_C - \mathbf{\Sigma}_C^N \textbf{G}_C^0$. Then its inverse is given by
\begin{equation}
\label{eq:BCInverse}
\textbf{B}_C^{-1} = \left[ \begin{array} {cc}
\frac{1 - \Sigma_{NN} G_{NN}^0} {\Lambda} & \frac{\Sigma_{11} G_{1N}^0} {\Lambda} \\
\frac{\Sigma_{NN} G_{N1}^0} {\Lambda} \;\; & \frac{1 - \Sigma_{11} G_{11}^0} {\Lambda}
\end{array} \right],
\end{equation}
where $\Lambda$ denotes the determinant of $\textbf{B}_C$ and is equal to
\begin{equation}
\label{eq:Lambda}
\Lambda = (1 - G_{11}^0 \Sigma_{11})(1 - G_{NN}^0 \Sigma_{NN}) - \Sigma_{11} \Sigma_{NN} G_{1N}^0 G_{N1}^0.
\end{equation}

The local density of states (LDOS) of 1D open systems, $\rho(z,E)$, takes the form of
\begin{equation}
\label{eq:rho1d}
\rho(z, E) = \frac{1}{2 \pi} ( |G_{z 1}^R|^2 \Gamma_{11} + |G_{z N}^R|^2 \Gamma_{NN} ).
\end{equation}
$G_{z m}^R \; (m = 1, N)$ is the retarded Green's function of an open device in a mixed space and mode representation \cite{Mamaluy2007} and is computed as
\begin{eqnarray}
\label{eq:GrmR}
G_{z m}^R &=& \langle z | \textbf{G}_C^0 \textbf{B}_C^{-1} | m \rangle \nonumber \\
&=& \sum_{m^\prime, \alpha} \frac{\langle z | \alpha \rangle \langle \alpha |m^\prime \rangle} {E -\varepsilon_\alpha} \langle m^\prime | \textbf{B}_C^{-1} | m \rangle \nonumber \\
&=& \sum_\alpha \frac{ \psi_\alpha(z) } {E - \varepsilon_\alpha} [\psi_\alpha^{*}(1) \textbf{B}_{1m}^{-1} + \psi_\alpha^*(N) \textbf{B}_{Nm}^{-1} ].
\end{eqnarray}
From the above equation and Eq.~(\ref{eq:BCInverse}), we can rewrite $G_{z m}^R$ as follows,
\begin{eqnarray}
\label{eq:xyalpha}
G_{z 1}^R &=& \sum_\alpha \psi_\alpha(z) x_\alpha(E), \nonumber \\
G_{z N}^R &=& \sum_\alpha \psi_\alpha(z) y_\alpha(E), \nonumber \\
x_\alpha(E) &=& \frac{\psi_\alpha^*(1) (1-\Sigma_{NN} G_{NN}^0) + \psi_\alpha^*(N) \Sigma_{NN} G_{N1}^0} {\Lambda (E - \varepsilon_\alpha)}, \nonumber \\
y_\alpha(E) &=& \frac{\psi_\alpha^*(1) \Sigma_{11} G_{1N}^0 + \psi_\alpha^*(N) (1 - \Sigma_{11} G_{11}^0) } {\Lambda (E - \varepsilon_\alpha)}.
\end{eqnarray}

Finally, the electron density $n(z)$ is computed as
\begin{equation}
\label{eq:quantumdensity}
n(z) = \int [\rho_1(z,E) f_1^{1D}(E) + \rho_N(z,E) f_N^{1D}(E)] dE,
\end{equation}
with
\begin{eqnarray}
\label{eq:rho1N_f1D1N}
\rho_1(z,E) &=& \frac{1} {2 \pi a} \biggl| \sum_\alpha \psi_\alpha(z) x_\alpha(E) \biggl|^2 \Gamma_{11}, \nonumber \\
\rho_N(z,E) &=& \frac{1} {2 \pi a} \biggl| \sum_\alpha \psi_\alpha(z) y_\alpha(E) \biggl|^2 \Gamma_{NN}, \nonumber \\
f_1^{1D}(E) &=& \frac{m_{\parallel}^* k_B T} {\pi \hbar^2} \ln \biggl[ 1 + \exp \biggl( \frac{E_{F1} - E}{k_B T} \biggl) \biggl], \nonumber \\
f_N^{1D}(E) &=& \frac{m_{\parallel}^* k_B T} {\pi \hbar^2} \ln \biggl[ 1 + \exp \biggl( \frac{E_{FN} - E}{k_B T} \biggl) \biggl],
\end{eqnarray}
where $1/a$ is added in LDOS to obtain the correct units for the electron density. $f_1^{1D}(E)$ and $f_N^{1D}(E)$ are the 1D distribution functions (in units of 1/cm$^2$ and including spin degeneracy) of the left and right leads respectively, obtained through integrating the Fermi-Dirac functions in the leads over an infinite 2D cross section \cite{DattaQuanTran}. $m_{\parallel}^*$ is the electron density-of-states effective mass in the 2D plane. $E_{F1}$ and $E_{FN}$ are the quasi-Fermi levels of the leads. If lead $1$ has zero voltage, and lead $N$ has a voltage of $V_{DS}$, then $E_{F1}$ = 0 and $E_{FN} = - q V_{DS}$.

The current in 1D devices is computed as
\begin{equation}
\label{eq:Current1D}
I^{1D} = \frac{q}{h} \int T^{1D}(E) [f_1^{1D}(E) - f_N^{1D}(E)] dE,
\end{equation}
which is in the units of Amperes/cm$^2$.

\section{Predictor-Corrector Scheme}
\label{appendix:selfconsistency}

Following the procedure in Ref.~\onlinecite{Mamaluy2007}, we first start the CBR transport assuming $\varphi = 0$ or using the solution of $\varphi$ at a previous voltage point, when voltage sweeping is enabled, to compute the electron density $n(z)$ using the expression in Eq.~(\ref{eq:quantumdensity}). The potential $\varphi$ and the electron density $n$ are then used to calculate the residuum $F$ of the Poisson equation using
\begin{equation}
\label{eq:PoFunctional}
F[\varphi] = \textbf{A} \varphi + (n - N_D),
\end{equation}
where $\textbf{A}$ is the matrix derived from the discretization of the Poisson equation, Eq.~(\ref{eq:Poisson}), with \emph{Neumann} BC. (Neumann BC is a more appropriate boundary condition as the electric field should be zero in a lead, while Dirichlet BC does not guarantee zero electric field.)

If the L1 norm $||F||$ of the residuum is smaller than a pre-defined threshold, the CBR-Poisson iteration is taken as converged. If the norm is still large, the correction to the Hartree potential, $\Delta \varphi(z)$, is obtained in the predictor step by solving the nonlinear Poisson equation below
\begin{eqnarray}
\label{eq:prPoisson}
&& \textbf{A} [\varphi_k^{in}(z) + \Delta \varphi(z)] = -n_{pr}(z, \Delta \varphi(z)) + N_D(z), \nonumber \\
&& n_{pr}(z, \Delta \varphi(z)) = \sum_{j=1,N} \int \rho_j(z,E) f_j^{1D}(E, \Delta \varphi(z)) dE, \nonumber \\
&& f_j^{1D} = \frac{m_{\parallel}^* k_B T} {\pi \hbar^2} \ln \biggl[ 1 + \exp \biggl( \frac{E_{Fj} - E + q \Delta \varphi(z) }{k_B T} \biggl) \biggl], \nonumber \\
\end{eqnarray}
where $\rho_j(z,E) \; (j = 1, N)$ is given in Eq.~(\ref{eq:rho1N_f1D1N}), $\varphi_k^{in}(z)$ is the input potential for the $k$th CBR-Poisson iteration and it is equal to the output potential of the previous $(k-1)$th iteration in the predictor-corrector scheme (so it is a known quantity). To solve the above nonlinear equation, we need to know the Newton Jacobian matrix, which can be found analytically as follows
\begin{eqnarray}
\label{eq:newtonJacobian}
\textbf{J}_{zz'} &=& \frac{\partial F_{pr}(z)} {\partial \Delta \varphi(z')} = \textbf{A}_{zz'} + \frac{\partial n_{pr}(z)} {\partial \Delta \varphi(z')} \nonumber \\
&=& \textbf{A}_{zz'} + \delta_{zz'} \sum_{j=1,N} \int \rho_j(z,E) \frac{\partial f_j^{1D}} {\partial \Delta \varphi(z')} dE, \nonumber \\
\frac{\partial f_j^{1D}} {\partial \Delta \varphi(z')} &=& \frac{q m_\parallel^*} {\pi \hbar^2} \frac{1} {1+\exp \biggl( \frac{E - q \Delta \varphi(z) - E_{Fj}} {k_B T} \biggl) }.
\end{eqnarray}

The Newton method used to solve Eq.~(\ref{eq:prPoisson}) consists of two main steps: (i) for the $(m+1)$th Newton iteration, compute the Jacobian matrix $\mathbf{J}$ using Eq.~(\ref{eq:newtonJacobian}), which is a tridiagonal matrix for 1D; (ii) solve the tridiagonal linear problem, $\mathbf{J} (\Delta \varphi_{m+1} - \Delta \varphi_{m}) = -F_{pr}$ to obtain $\Delta \varphi_{m+1}$, where $F_{pr} = \textbf{A} [\varphi_k^{in} + \Delta \varphi_{m}] + n_{pr}(\Delta \varphi_{m}) - N_D$, and $\Delta \varphi_{m}$ is the $m$th Newton solution. When $||\Delta \varphi_{m+1} - \Delta \varphi_{m}||$ is less than a pre-defined small value, e.g., $10^{-13}$ used in the simulation, $\Delta \varphi_{m+1}$ is taken to be the solution $\Delta \varphi$ of the nonlinear Eq.~(\ref{eq:prPoisson}). It is worthy of noting that we use the Newton method with a line search algorithm \cite{NumericRecipes}, which has shown robust convergence for the nonlinear Poisson equation considered here [cf. Eq.~(\ref{eq:prPoisson})].

The potential correction $\Delta \varphi$ obtained from the Newton method is used to update the output potential at the $k$th CBR-Poisson iteration
\begin{equation}
\varphi_k^{out} = \varphi_k^{in} + \Delta \varphi.
\end{equation}
The input potential for the next $(k+1)$th iteration, $\varphi^{in}_{k+1}$, is set equal to $\varphi_k^{out}$, i.e., $\varphi^{in}_{k+1} = \varphi_{k}^{out}$. Then the CBR-Poisson loop is repeated until convergence is achieved, at which the Poisson residuum $|| F[\varphi] || < \varepsilon$ with $\varepsilon$ being a user-defined threshold, and $|| \Delta \varphi ||$ is numerically close to 0, implying that the difference between the predictor electron density $n_{pr}(z)$ and the quantum electron density $n(z)$ is negligible.

\section{3D Current}
\label{appendix:3Dcurrent}

We know the current for a 1D device with a uniform cross section is in units of Amperes/cm$^2$ and computed using Eq.~(\ref{eq:Current1D}). To obtain the total current in Amperes, we need to know the area of the cross section, i.e., $I^{3D} = I^{1D} A$, with $A$ being the cross-section area. On the other hand, the actual tunnel barrier in a DQD is not 1D, but 3D in nature, implying that in principle, we need a 3D quantum transport simulation. The actual current in Amperes should be computed as
\begin{equation}
\label{eq:Current3D}
I^{3D} = \frac{2 q} {h} \int T^{3D} (f_L^{3D} - f_R^{3D}) dE,
\end{equation}
where 2 is the spin double degeneracy and $T^{3D}$ is the 3D transmission function. $f_L^{3D}$ and $f_R^{3D}$ are the Fermi-Dirac distributions of the left and right leads, respectively.

To compute the 3D current from the self-consistent 1D results, we need to satisfy the following relation
\begin{eqnarray}
I^{3D} &=& A \frac{q}{h} \int T^{1D} (f_1^{1D} - f_N^{1D}) dE  \nonumber \\
&=& \frac{2q}{h} \int T^{3D} (f_L^{3D} - f_R^{3D}) dE,
\end{eqnarray}
which implies
\begin{equation}
T^{3D}(E) = \frac{1}{2} A T^{1D} \frac{f_1^{1D} - f_N^{1D}} {f_L^{3D} - f_R^{3D}} = T^{1D}(E) \gamma(E),
\end{equation}
where $\gamma(E)$ is an unknown energy-dependent function. To obtain an approximate $T^{3D}(E)$, we neglect the energy dependence of $\gamma$, and compute $I^{3D}$ as
\begin{equation}
\label{eq:approxI3D}
I^{3D} = \gamma \frac{2q}{h} \int T^{1D} (f_L^{3D} - f_R^{3D}) dE.
\end{equation}
where $\gamma$ is determined by matching the computed drain-source differential conductance $G_{DS}$ to the right order of magnitude when compared to the measured differential conductance. $G_{DS}$ is calculated from the $I^{3D}$-V relation using the finite center difference method.

\end{document}